# General Classification for Digital Knowledge: Algorithm, Identifiers and a First Benchmark

David Awad

**Abstract**

Digital libraries and knowledge graphs need a way to organize very large collections so that related works are close and any work is quick to reach. The classifications libraries inherited for this, such as LCC and DDC, are maintained by editorial staff, are partly closed, and were not built against a stated, measurable objective. We propose the opposite: a general classification that is *fitted*, under the constraints that make a classification usable on a shelf or in a browse list, to the same quantity that a companion survey and benchmark [1] ranks existing schemes by. The objective is $J = D + 0.005N$, where $D$ is the normalized Dasgupta cost of the class tree on a fixed similarity graph [2] and $N$ is an expected navigation cost in bits with a per-level charge. Two constraints are hard: every class is a contiguous run of one fixed shelf order (the optimal-alphabetic-tree constraint of Hu and Tucker [3]), and new works are filed by splitting leaf classes only, so that every existing class keeps its code, label and parent. We describe a deterministic fitting algorithm (average linkage, rotation local search, an exact dynamic-programming collapse to 2–12 children and at most 4 levels, and order-preserving local search), state the identifier-stability guarantee precisely, and tie each clause of it to a test. On a corpus of 10,000 arXiv papers the fitted prototype has 110 hand-labelled classes and reaches cost 0.1722 in sample (average linkage 0.1518; arXiv's own categories 0.3333; the best existing scheme reached through the same lexical classifier 0.6414, from the survey). Fitted on 80% and tested on the other 20%, it scores 0.1722. Filing 500 library books as new works created 16 classes and changed no existing identifier. Rescored on graphs from two independent embedding models, the scheme still leads every existing scheme in all 9 graph conditions, with a smaller margin. Under a corrected objective that also charges for joining dissimilar works, it still leads on the papers, but on 3,000 library books its lead over librarian-assigned Dewey and LCC numbers depends on the graph setting. These numbers flatter the method. Fit, filing and scoring share one embedding: the held-out skill of 0.922 falls to 0.896 when the same fit is scored on the graphs of the two other models, and on books the average falls from 0.806 to 0.731. The filing step, which always uses the fit's own embedding, is not separated by any run. The scheme has also seen only arXiv, and labels for inserted classes are weak. We state those limits plainly, release the scheme as CC0 open data, and ask whether a learned, open, order-preserving classification can become a general one.

**Keywords:** library classification; hierarchical clustering; Dasgupta cost; order-preserving trees; identifier stability; open data

## 1 INTRODUCTION

### 1.1 Why a learned general classification

A general classification gives every work one address in one linear order. In a digital library or a knowledge graph the collection is large, grows continuously, and is filed by software, so the cost of finding and filing a work is what an organization has to keep low. That order serves two readers: the person browsing a shelf or a browse list, who wants related works next to each other, and the cataloguer or system that must file each new work somewhere. The schemes libraries use for this, LCC and DDC above all, were enumerated by hand and have been extended editorially since DDC's first edition of 1876 [4], [5] and LCC's early twentieth-century schedules [6]. Open universal alternatives exist, but none is both deep and open with a full crosswalk, and none of them, open or closed, was designed against a stated, measurable objective, so there is no way to say whether one is better than another for a given collection except by argument. The survey of those schemes, with their licences, is in the companion survey and is not repeated here.

We previously measured 13 published schemes [1] on one objective scale: the Dasgupta cost of each scheme's class tree over a fixed kNN similarity graph of 10,000 arXiv papers, with navigation statistics alongside. That benchmark makes the converse question natural. If a scheme can be scored, can a scheme be *fitted* to the score, under the constraints a library needs, and how far does it get? This paper is that proposal.

Data-driven classifications already exist at scale. The CWTS publication-level classification clusters a citation network by a modularity-type function at several resolutions [7], [8], and OpenAlex Topics assigns works to citation-derived topics whose upper levels are mapped from an existing journal classification [9], [10]. Topic-taxonomy builders such as TaxoGen [11] and LLM-driven pipelines such as Chain-of-Layer and TnT-LLM [12], [13] produce trees from text. None of these optimizes a shelf order, and none gives identifiers a guarantee under growth. Those two properties are what separates a classification from a clustering for a library.

### 1.2 Why order preservation matters

A shelf, a call-number index, and a browse list are each one linear order. A hierarchy is useful for them only if every class is a contiguous run of that order, so that "go to class X" means "go to one place". A tree fitted without this constraint can always be drawn in *some* leaf order, but a search that freely regroups works across the order produces classes whose works are not adjacent once the order is fixed. We therefore fix the shelf order first and restrict every later step to trees whose leaf order is that shelf order. This is the alphabetic-tree constraint of optimal order-preserving codes and search trees [3], [14], as opposed to the free leaf order of Huffman coding [15].

### 1.3 Why identifier stability matters

Class codes are written into call numbers, spine labels, catalogue records, crosswalks and links. A classification that is refitted from scratch each year would renumber works that never moved, invalidating all of those. For a learned scheme this is the decisive practical objection: an optimizer will happily trade any existing code for a small gain in the objective. We therefore separate *fitting* (done once, in batch) from *growth* (done incrementally), and constrain growth so that existing identifiers are never changed, only refined. TaxoExpan [16] inserts new nodes into a frozen taxonomy, but we found no objective in the literature that trades optimality against re-shelving.

### 1.4 Contributions

1. An objective for a general classification that combines similarity coherence and navigation cost, with explicit shape constraints, computed by the same harness that scores published schemes (Section 2).
2. A deterministic, order-preserving fitting algorithm, with an exact dynamic program for the collapse of a binary tree to a bounded-shape class tree (Section 3).
3. An insertion procedure for new works with a precisely stated identifier-stability guarantee, checked on every run and by tests on real data (Section 4).
4. A hand-labelled prototype, `scheme:learned@2026-10-10`, released under CC0, and its scores against the 13 published schemes of the survey, in sample, through a lexical classifier, on a held-out split, and on a holdout of library books (Sections 5 to 6 and 9). The embedding circularity of these scores is quantified in Section 8.1.

## 2 OBJECTIVE

Let $G$ be a weighted similarity graph over $n$ works, with edge weights $w_{ij}$ and total weight $W$. Here $G$ is the graph published with the survey: the sentence-transformers/all-MiniLM-L6-v2 embedding [17], [18], [19] of title and abstract, cosine similarity minus the mean pair cosine, $k = 20$ nearest neighbours, symmetrized by maximum (148,111 edges over $n = 10{,}000$ works). A class tree $T$ places each work in one leaf class; a virtual root holds the top-level classes.

**Similarity term.** Works hang under their leaf class, so two works in one class meet at that class and pay its size:

$$D(T) = \frac{1}{nW} \sum_{(i,j)\in E} w_{ij} |\text{items}(\text{lca}_T(i,j))|. \qquad (1)$$

This is Dasgupta's cost [2], normalized so that a flat tree scores 1, and it is exactly the harness's `cost_norm`. Its complement is the Moseley–Wang revenue [20]. Exact minimization is hard; the approximation and axiomatic literature [21], [22] and continuous relaxations [23] concern binary trees without shape or order constraints, which is why we start from average linkage and add the constraints in later steps. The survey's harness also reports skill $= \frac{1-D}{1-D_{\text{ref}}}$, where $D_{\text{ref}}$ is the cost of the average-linkage tree on the same graph; this paper uses it as the normalised score.

**Navigation term.** The fan-out of a class is its number of child classes, or its number of works if it is a leaf class. Then

$$N(T) = \frac{1}{n} \sum_{v:\ \text{fanout}(v)>1} |\text{items}(v)|(a + \log_2 \text{fanout}(v)) \quad (2)$$

is the mean, over works, of the cost of descending from the root to the work, charging $a$ per menu level plus $\log_2$ of the menu size, a Hick–Hyman-style choice cost [24], [25]. It is the harness's `nav_log_a1`. A balanced tree pays about $\log_2 n$ bits plus $a$ per level, so $a$ prices a level and the $\log_2$ part penalizes fan-outs that do not match the sizes below them. Without $a$, depth would be free; without the log term, the cheapest tree would be flat.

**Objective and constraints.**

$$J(T) = D(T) + \lambda N(T), \quad \lambda = 0.005, \quad a = 1 \qquad (3)$$

subject to: every class has 2–12 child classes, at most 4 class levels, leaf classes of at least 5 works, and every class a contiguous run of one fixed shelf order with siblings ordered along it. With these weights, one extra level for every work is worth a 0.005 drop in $D$: 0.5% of the corpus scanned per related pair.

The two terms pull against each other. $D$ falls with every split and is smallest for a binary tree with one work per leaf; $N$ charges $a$ bits per level to every work under it. The choice of $\lambda$, $a$ and the depth bound is a modelling decision, not a fitted quantity; Section 6.5 reports a sweep. We do not claim that $J$ is *the* objective for a general classification: the optimal tree is defined relative to the similarity $w$, and there are formal reasons to doubt any similarity-free optimum [26], [27]. The claim is narrower: once a benchmark (the survey) ranks schemes by $D$ and $N$, a scheme can be fitted to them, and the result is a reference point for every hand-built scheme on the same scale.

## 3 ALGORITHM

`learned-scheme build` (`schemes/learned/src/fit.rs`) is deterministic for a fixed seed (20261010, used only for local-search visiting orders).

1. **Average linkage on the sparse graph** (`src/binary.rs`). Cluster similarity is the summed edge weight over the product of the cluster sizes, missing pairs counting 0; this is Moseley and Wang's average linkage, which approximates the optimal Dasgupta revenue within a factor of 1/3 [20]. Ties go to the lowest cluster ids; disconnected components are joined smallest first. The resulting binary tree's cost, 0.151834, equals the harness's reference value.

**Table 1.** Fitting stages on the 10,000-paper corpus ($D$ is `cost_norm`). The four stages take 2.2 s of wall time in total on one core.

| stage | $D$ | $N$ $(a=1)$ | $J$ | classes | s |
|---|---|---|---|---|---|
| average linkage (binary, one work per leaf) | 0.151834 | - | - | - | 0.91 |
| rotation local search (0 moves) | 0.151834 | - | - | - | 0.04 |
| exact collapse (DP) | 0.172309 | 18.4841 | 0.264730 | 129 | 0.05 |
| order-preserving local search (88 shifts, 20 structure moves) | 0.172156 | 18.4487 | 0.264399 | 110 | 1.19 |

2. **Rotation local search** on $D$. At a node $x = (A, C)$ with $A = (A_1, A_2)$, try $((A_1, C), A_2)$ and $((A_2, C), A_1)$ with exact local deltas. On this graph it makes 0 moves. That is expected: a rotation helps exactly when the average weight between $A_1$ and $C$ exceeds that between $A_1$ and $A_2$, and average linkage is reducible, so when $A_1$ and $A_2$ merged no part of a later $C$ was closer. The pass certifies rotation-optimality and matters for other starting trees.
3. **Shelf order.** Each merge's two children are oriented so that the child more strongly tied to the works already shelved to its left comes first. The binary tree's leaf order becomes the fixed shelf order; nothing later changes it.
4. **Exact collapse** (`src/collapse.rs`). A dynamic program over the binary tree chooses which binary nodes become classes, contracting the rest, to minimize $J$ over all such class trees within the shape bounds. For a class $u$ it chooses the frontier of descendant classes by a knapsack over the binary nodes in between. Binary subtrees with fewer than 5 works are glued to their sibling and join the adjacent leaf class on the shelf. The DP is exact when glue sits only inside leaf classes (proved in Lean). Above leaf level it has no guarantee: small counterexamples show its value can be either too high or too low. Here it is close ($J$ 0.264697 by the DP, 0.264730 evaluated exactly). Every tree this step can produce is order-preserving by construction, because contracting binary nodes never reorders leaves.
5. **Order-preserving local search** (`src/search.rs`) on exact $J$: shift a boundary work into the neighbouring leaf class on the shelf (incremental delta), hand the last child class to the next sibling or the reverse, merge adjacent sibling leaves, flatten a class into its parent, group two or three adjacent siblings under a new class. Every move keeps the shape bounds and the shelf order. Rounds repeat until no move improves $J$.

Table 1 gives the stages on the corpus (from `build-report.json`).

The result has 110 classes (2 / 5 / 16 / 87 by level), 88 of them leaf classes of 38–345 works (median 98). Class codes are the parent's code plus a two-digit position (`01`, `01.02`, `01.02.03`), so codes sort in shelf order. The root has 2 children; internal classes have at most 8 (mean 4.8). Table 2 shows the top three levels.

The test `rebuild_reproduces_the_committed_scheme_and_matches_the_harness` (`schemes/learned/tests/committed.rs`) rebuilds the scheme byte for byte from committed inputs and checks that $D$ and $N$ as computed by the builder equal the harness's `cost_norm` and `nav_log_a1` for the committed scheme. `make_tables.py` repeats the second check on the published numbers.

**Table 2.** The top three levels of `scheme:learned@2026-10-10`, with the number of leaf classes and fitted works below each class.

| code | label (hand-written) | leaf classes | works |
|---|---|---|---|
| `01` | Computing, AI and engineering | 50 | 5,305 |
| `01.01` | Language models, security and computing systems | 18 | 2,555 |
| `01.01.01` | Language, speech and LLM agents | 5 | 1,221 |
| `01.01.02` | Security, privacy and software | 6 | 581 |
| `01.01.03` | Efficient computing, networks and energy systems | 7 | 753 |
| `01.02` | Vision, robotics, dynamics and medicine | 22 | 1,927 |
| `01.02.01` | Computer vision and multimodal models | 4 | 585 |
| `01.02.02` | Robotics and control | 5 | 424 |
| `01.02.03` | Machine learning for physical and dynamical systems | 8 | 371 |
| `01.02.04` | Medical imaging and health | 5 | 547 |
| `01.03` | Statistics, economics and decision-making | 10 | 823 |
| `01.03.01` | Statistical and causal inference | 5 | 290 |
| `01.03.02` | Financial forecasting and markets | 1 | 216 |
| `01.03.03` | Decision-making, learning and games | 4 | 317 |
| `02` | Mathematics and physical sciences | 38 | 4,695 |
| `02.01` | Mathematics | 14 | 2,251 |
| `02.01.01` | Applied mathematics, analysis and probability | 7 | 1,105 |
| `02.01.02` | Pure mathematics | 7 | 1,146 |
| `02.02` | Physics | 24 | 2,444 |
| `02.02.01` | Quantum information and quantum computing | 6 | 465 |
| `02.02.02` | Condensed matter, materials and chemistry | 7 | 836 |
| `02.02.03` | Astrophysics and particle physics | 6 | 808 |
| `02.02.04` | Gravitation and theoretical physics | 5 | 335 |

## 4 IDENTIFIER STABILITY UNDER INSERTION

### 4.1 Procedure

`learned-scheme insert` (`src/insert.rs`) files new works one at a time without renaming anything.

1. A new work goes to the leaf class whose centroid is nearest (cosine on unit sentence-transformers/all-MiniLM-L6-v2 vectors; the committed centroids are 384-dimensional, i8-quantized, one per leaf class with its work count).
2. A leaf class tries a two-means split, anchored on its fitted centroid, when either (a) it has taken at least 12 new works and they form a distinct group (the two centroids' cosine is at most 0.5), or (b) its new works reach its capacity: its fitted size, or the median fitted leaf size (98) for a class created by insertion. Capacity splits start from the first principal direction of the new works [28]. Both sides need at least 5 works.

3. A split turns leaf $C$ into the parent of two new classes: `C.01`, which keeps the core ($C$'s earlier works and the new works near them) and is labelled "<$C$'s label> (general)", and `C.02`, which takes the other group and is labelled from its works' most frequent title words.

## 4.2 Guarantee

Let $S$ be a scheme and $S'$ the scheme after inserting any sequence of works. Then:

- **(G1) Existing classes are unchanged.** Every class of $S$ is in $S'$ with the same code, label, vocabulary, note and parent, and the classes of $S$ appear in $S'$ in the same relative order.
- **(G2) Only leaves are refined.** Every class of $S'$ that is not in $S$ has a parent, that parent was a leaf class of $S$ or is itself a class created by insertion, and the new class's code extends its parent's code (`C` to `C.01`, `C.02`).
- **(G3) Existing works only move down.** A work filed in class $C$ of $S$ is in $C$ or in a descendant of $C$ in $S'$; when $C$ is split, its earlier works go to `C.01`.
- **(G4) Shelf positions of existing works are kept.** Because `C.01` opens $C$'s run and `C.02` follows it, a work that was in $C$ keeps its position relative to every work of $S$.

G1 and the parent clause of G2 are what `insert::check_stable` checks; `learned-scheme insert` runs it on every insertion and writes no output scheme if it fails. The test `holdout_insertion_keeps_every_existing_id_and_parent_link` (`schemes/learned/tests/insertion.rs`) inserts the 500-book holdout and checks G1 class by class (code, parent, label, vocabulary, relative order), G2 including the code-extension clause, that exactly two classes appear per split, that every new work sits in a current leaf class, and the G3 clause that a split leaf's earlier works move only to a descendant. `insertion_is_deterministic` checks that the same inputs give byte-identical output, and `stability_check_rejects_changes_to_existing_classes` (`tests/insert_rules.rs`) checks that `check_stable` rejects a relabelled class, a new class under an existing internal class, and a reordering. G4 follows from the code grammar and the core-first construction; it is not separately tested beyond G3.

The consequence for a catalogue: $S'$ is a conservative extension of $S$. Its codes and parent links are a superset of those of $S$, an existing call number stays valid (it may become less specific than it could be, never wrong), and the inserted scheme takes the version `<version>.ins<count>`, so a client that compares versions sees only the split leaves as changed.

What the guarantee does *not* say: it bounds nothing about $J$ after insertion, it does not keep the depth bound (a leaf can be split repeatedly), and it says nothing about label quality for the new classes. Section 8 returns to each of these.

# 5 LABELS

No language model and no API was used. The 110 fitted classes were labelled by hand, by the AI coding assistant that built the prototype, from a label sheet (`learned-scheme build --sheet`). For every class the sheet shows its size, its top c-TF-IDF terms, and its most central titles (largest summed graph weight to the class's other works). c-TF-IDF is class-based TF-IDF as in BERTopic [29], scoring term $t$ in class $c$ by $\frac{\mathrm{tf}(t,c)}{\mathrm{tokens}(c)} \cdot \ln\left(1 + \frac{A}{f(t)}\right)$ over titles counted twice and abstracts, unigrams and adjacent bigrams, keeping terms present in a minimum share of the class's works (`src/terms.rs`). The top terms also become the class's vocabulary in the scheme file (source `ctfidf`), which is how shelfmark's lexical classifier reaches the class.

Each label line in `schemes/learned/labels.tsv` stores the class's top terms as a fingerprint, and the builder refuses a label whose fingerprint no longer matches, so a refit that moves classes cannot keep stale labels silently. `committed_scheme_is_valid_and_every_label_is_hand_written` checks that no class of the committed scheme carries an automatic label. Classes created by insertion get automatic labels (Section 4.1), and the 80% fit used for the held-out split uses automatic labels throughout.

We note the distinction from LLM taxonomy pipelines [12], [13]: there the model proposes the structure; here the structure is fixed by the objective and labelling is a separate, auditable step. Whether hand labels written this way are good labels is a question for human evaluation that this draft does not answer (Section 8).

# 6 RESULTS

All runs use the survey's corpus (10,000 arXiv papers, `corpus.jsonl` sha256 `7685685d0de2`...), holdout (500 Open Library editions with LC call numbers, `holdout.jsonl` sha256 `3e057eb4a257`...), published $k = 20$ graphs (corpus graph sha256 `759d3b8461a9`...) and harness; The survey describes their construction and validity. Its 19 placed and 5 gold scores were regenerated with `profile:lexical@2` and equal the published ones (checked in `results.json` and again by `make_tables.py`). Lower is better for `cost_norm`, $J$ and $N$; higher is better for skill. Three kinds of placement appear: *gold* (a source's own labels, e.g. arXiv categories or LC call numbers), *shelfmark lexical@2* (a scheme reached through the survey's lexical classifier, using only labels and vocabulary), and *own* (the learned scheme's fitted assignment or its insertion procedure). Unplaced works hang under the root, so low coverage raises cost. The tables below keep the learned rows and the comparison rows the text cites; every row is in `bench/results/2026-10-10-learned/results.json`, and the survey reports the other schemes' results and their dependence on coverage.

## 6.1 In sample (the corpus it was fitted to)

In sample the learned scheme reaches cost 0.1722, skill 0.9760: 97.6% of the average-linkage revenue with 110 named classes and 4 levels. arXiv's own author-chosen categories score 0.3333 ($J$ 0.4250). The best published scheme reached through the lexical classifier is openalex-topics at 0.6414; the worst corpus-wide placed row is msc at 0.9498 (the survey explains why placed rankings track coverage). This row is the objective the scheme was fitted to, evaluated on the data it was fitted on, so it is an upper bound on what the method shows, not evidence of generalization.

**Table 3.** Corpus, $n = 10{,}000$, learned rows against the arXiv gold row and the best and worst surveyed scheme placed by the lexical classifier. The first learned row is in sample by construction.

| scheme | placements | coverage | $D$ | skill | $J$ | $N$ $(a = 1)$ | classes used | mean depth | max branching |
|---|---|---|---|---|---|---|---|---|---|
| `learned` | own, in-sample (fitted on this corpus) | 1.0000 | 0.1722 | 0.9760 | 0.2644 | 18.45 | 39.7 | 4.98 | 345 |
| `arxiv` | gold (source labels) | 1.0000 | 0.3333 | 0.7861 | 0.4250 | 18.33 | 36.5 | 3.90 | 797 |
| `learned` | shelfmark lexical@2 | 0.9993 | 0.3794 | 0.7317 | 0.5347 | 31.08 | 33.5 | 4.38 | 1183 |
| `openalex-topics` | shelfmark lexical@2 | 0.9732 | 0.6414 | 0.4228 | 0.7757 | 26.85 | 77.6 | 4.45 | 554 |
| `msc` | shelfmark lexical@2 | 0.3916 | 0.9498 | 0.0592 | 1.0389 | 17.82 | 114.4 | 2.00 | 380 |
| `average linkage (binary reference, not a classification)` | - | 1.0000 | 0.1518 | 1.0000 | - | - | - | - | 2 |

**Table 4.** Held-out test split, $n = 2{,}000$, learned rows against the arXiv gold row and the best surveyed scheme placed by the lexical classifier.

| scheme | placements | coverage | $D$ | skill | $J$ | $N$ $(a = 1)$ | classes used | mean depth | max branching |
|---|---|---|---|---|---|---|---|---|---|
| `learned (fit on the other 80%)` | own insertion, held out | 1.0000 | 0.1722 | 0.9222 | 0.2528 | 16.10 | 33.9 | 5.00 | 69 |
| `learned (full fit)` | own, in-sample | 1.0000 | 0.1793 | 0.9144 | 0.2600 | 16.15 | 39.2 | 4.98 | 71 |
| `arxiv` | gold (source labels) | 1.0000 | 0.3424 | 0.7327 | 0.4220 | 15.92 | 35.7 | 3.90 | 167 |
| `learned (full fit)` | shelfmark lexical@2 | 0.9995 | 0.3798 | 0.6910 | 0.4946 | 22.98 | 33.0 | 4.39 | 238 |
| `openalex-topics` | shelfmark lexical@2 | 0.9750 | 0.6519 | 0.3878 | 0.7580 | 21.22 | 70.7 | 4.43 | 119 |

**Table 5.** The 80% fit in sample, the 80% fit with the 20% inserted, and the full fit, on the corpus graph.

| scheme | placements | coverage | $D$ | skill | $J$ | $N$ $(a = 1)$ | classes used | mean depth | max branching |
|---|---|---|---|---|---|---|---|---|---|
| `learned (fit on 80%)` | own, in-sample, 80% graph | 1.0000 | 0.1665 | 0.9777 | 0.2570 | 18.11 | 33.7 | 5.00 | 259 |
| `learned (fit on 80%)` | 80% fitted + 20% inserted, whole corpus graph | 1.0000 | 0.1720 | 0.9763 | 0.2641 | 18.42 | 33.8 | 5.00 | 328 |
| `learned (full fit)` | own, in-sample, whole corpus graph | 1.0000 | 0.1722 | 0.9760 | 0.2644 | 18.45 | 39.7 | 4.98 | 345 |

**Table 6.** Holdout, $n = 500$ books, learned rows against LCC gold rows and the best surveyed scheme placed by the lexical classifier.

| scheme | placements | coverage | $D$ | skill | $J$ | $N$ $(a = 1)$ | classes used | mean depth | max branching |
|---|---|---|---|---|---|---|---|---|---|
| `learned` | own insertion (embedding nearest centroid) | 1.0000 | 0.4432 | 0.7971 | 0.5416 | 19.69 | 15.1 | 8.79 | 93 |
| `lcc` | gold (source labels) | 0.9220 | 0.7076 | 0.4185 | 0.8006 | 18.59 | 35.0 | 4.20 | 89 |
| `lcc-full` | gold (source labels) | 0.9220 | 0.7057 | 0.4212 | 0.8043 | 19.71 | 85.2 | 5.19 | 68 |
| `lcc-full` | shelfmark lexical@2 | 0.5660 | 0.9250 | 0.1074 | 0.9949 | 13.98 | 340.8 | 4.14 | 13 |
| `learned` | shelfmark lexical@2 | 0.0860 | 0.9915 | 0.0122 | 1.0443 | 10.57 | 26.1 | 1.34 | 5 |
| `average linkage (binary reference, not a classification)` | - | 1.0000 | 0.3014 | 1.0000 | - | - | - | - | 2 |

## 6.2 Shelfmark-placed

The second learned row is the scheme reached by the survey's lexical classifier: cost 0.3794, $J$ 0.5347, coverage 99.93%. This removes the embedding from placement but not the fit: the classes are cut along the corpus graph's clusters and their vocabulary is the corpus's own most distinctive terms, so it is still in sample. It pays a higher navigation cost (31.08 bits against 18.45): the classifier concentrates works in fewer, larger classes (effective classes 33.5 against 39.7; largest fan-out 1183 against 345).

## 6.3 Held out: fitted on 80%, tested on 20%

The split ranks ids by SHA-256 with seed `learned-split-2026-10-10` the 2,000 test ids are committed in `bench/results/2026-10-10-learned/split/`. The 80% fit uses the same parameters with automatic labels and has 86 classes. The test works are then inserted as new works by nearest leaf centroid (0 splits occurred, so the scheme stayed at 86 classes). Everything in Table 4 is scored on the graph induced on the test works (average linkage 0.1024); every published scheme is restricted to the same works.

Held out, the learned scheme scores 0.1722 ($J$ 0.2528). The full fit, which saw these works, scores 0.1793 on them, so there is no sign of overfitting at this scale. arXiv's categories score 0.3424; the best published scheme through the classifier, openalex-topics, 0.6519. Table 5 compares the fits on the whole graph.

With the test works inserted, the 80% fit scores 0.1720 ($J$ 0.2641) on the whole corpus graph, against 0.1722 ($J$ 0.2644) for the full fit. Insertion lost nothing measurable here. The caveat that matters is in Section 8.1: insertion files works by the same embedding the graph is built from, so the held-out skill 0.9222 is optimistic.

## 6.4 Holdout: library books inserted as new works

The holdout is 500 Open Library editions with librarian-assigned LC call numbers, titles only, on its own $k = 20$ graph (average linkage 0.3014).

The books were inserted into the fitted scheme. The 110 classes became 126 through 8 leaf splits (3 for a distinct group, 5 for capacity), and every existing code, label and parent link was unchanged (Section 4.2). Table 7 lists the splits.

Because the scheme was fitted on arXiv only, most books first land in one fitted leaf, `01.02.01.04` "Image generation and editing", the nearest centroid to short book titles; the first split, after 45 books, separates them from that class's papers, and later capacity splits divide them further. On the holdout graph the inserted scheme scores 0.4432 (skill 0.7971), below the librarians' call numbers (LCC gold 0.7076 at 92.2% coverage, full LCC schedules gold 0.7057). Through the lexical classifier the learned scheme places only 43 of the 500 books (8.6%) and scores 0.9915; the best placed scheme on the holdout is lcc-full at 0.9250 (56.6% coverage). We do not read the 0.4432 as a win over LCC; Section 8.1 explains why.

**Table 7.** Leaf splits while inserting the holdout (`insert-report.json`). Split leaves among the fitted classes: `01.01.01.04` (Language technology, society and bias), `01.02.01.04` (Image generation and editing).

| after insert no. | cause | split leaf | new class | its label (automatic) | new works in it | centroid cosine |
|---|---|---|---|---|---|---|
| 45 | distinct | 01.02.01.04 | 01.02.01.04.02 | Harry / potter / adventures | 12 | 0.205 |
| 70 | distinct | 01.02.01.04.02 | 01.02.01.04.02.02 | Achterhuis / cak‘ / chan‘" | 5 | 0.304 |
| 158 | capacity | 01.02.01.04.02.01 | 01.02.01.04.02.01.02 | Court / man / power | 53 | 0.780 |
| 244 | capacity | 01.02.01.04.02.01.02 | 01.02.01.04.02.01.02.02 | Court / dog / power | 42 | 0.788 |
| 355 | distinct | 01.01.01.04 | 01.01.01.04.02 | Amigos / como / dick | 8 | 0.262 |
| 392 | capacity | 01.02.01.04.02.01.02.01 | 01.02.01.04.02.01.02.01.02 | Bell / beloved / blue | 48 | 0.801 |
| 397 | capacity | 01.02.01.04.02.01.01 | 01.02.01.04.02.01.01.02 | Edition / holmes / sherlock | 54 | 0.850 |
| 462 | capacity | 01.02.01.04.02.01.02.02 | 01.02.01.04.02.01.02.02.02 | Blood / body / cold | 54 | 0.773 |

**Table 8.** Sensitivity of the fit to $\lambda$ and the depth bound, $a = 1$ ($D$ is `cost_norm`). The committed scheme is depth bound 4, $\lambda = 0.005$.

| depth bound | $\lambda$ | $J$ | $D$ | $N$ $(a = 1)$ | classes | leaf classes | root children | median leaf |
|---|---|---|---|---|---|---|---|---|
| 3 | 0.001 | 0.2049 | 0.1875 | 17.46 | 59 | 50 | 2 | 174 |
| 3 | 0.002 | 0.2224 | 0.1875 | 17.46 | 59 | 50 | 2 | 174 |
| 3 | 0.005 | 0.2747 | 0.1877 | 17.41 | 54 | 45 | 2 | 226 |
| 3 | 0.01 | 0.3619 | 0.1879 | 17.40 | 51 | 42 | 2 | 247 |
| 3 | 0.02 | 0.5361 | 0.1881 | 17.40 | 49 | 40 | 2 | 262 |
| 3 | 0.05 | 1.0513 | 0.2075 | 16.88 | 33 | 28 | 3 | 307 |
| 4 | 0.001 | 0.1902 | 0.1717 | 18.50 | 125 | 100 | 2 | 81 |
| 4 | 0.002 | 0.2087 | 0.1717 | 18.49 | 120 | 95 | 2 | 87 |
| 4 | 0.005 | 0.2644 | 0.1722 | 18.45 | 110 | 88 | 2 | 98 |
| 4 | 0.01 | 0.3565 | 0.1731 | 18.33 | 92 | 73 | 2 | 121 |
| 4 | 0.02 | 0.5359 | 0.1860 | 17.50 | 52 | 42 | 2 | 243 |
| 4 | 0.05 | 1.0513 | 0.2075 | 16.88 | 33 | 28 | 3 | 307 |
| 6 | 0.001 | 0.1783 | 0.1577 | 20.68 | 330 | 248 | 2 | 31 |
| 6 | 0.002 | 0.1990 | 0.1581 | 20.44 | 271 | 204 | 2 | 45 |
| 6 | 0.005 | 0.2594 | 0.1600 | 19.90 | 207 | 161 | 2 | 56 |
| 6 | 0.01 | 0.3550 | 0.1675 | 18.76 | 101 | 77 | 2 | 119 |
| 6 | 0.02 | 0.5359 | 0.1860 | 17.50 | 52 | 42 | 2 | 243 |
| 6 | 0.05 | 1.0513 | 0.2075 | 16.88 | 33 | 28 | 3 | 307 |

**Table 9.** The learned scheme against the best other row in each of 9 graph conditions (three models, $k = 10, 20, 50$). Skill margin is learned skill minus the best other skill; $J$ margin is the best other $J$ minus learned $J$, so positive means the learned scheme is ahead on both. $J$ here is the fitted objective, `cost_norm` $+0.005N$.

| comparison | learned first by skill | skill margin CI $> 0$ | J margin CI $> 0$ | tightest condition: skill margin [95% CI] |
|---|---|---|---|---|
| held out 20% (MiniLM fit, inserted) | 9 of 9 | 9 of 9 | 9 of 9 | bge-small-k10: 0.155 [0.122, 0.189] |
| in sample (full MiniLM fit) | 9 of 9 | 9 of 9 | 9 of 9 | bge-small-k50: 0.160 [0.151, 0.169] |
| 500 holdout books, inserted | 9 of 9 | 9 of 9 | 9 of 9 | nomic-k10: 0.280 [0.214, 0.340] |
| placed by lexical@2 | 9 of 9 | 9 of 9 | 9 of 9 | nomic-k10: 0.290 [0.277, 0.302] |

### 6.5 Sensitivity

36 fits of the full corpus with automatic labels cover depth bounds 3, 4, 6, $\lambda$ in 0.001, 0.002, 0.005, 0.01, 0.02, 0.05 and $a$ in 1, 2 (`results.json`, `sweep`). Table 8 shows $a = 1$.

Raising $\lambda$ trades $D$ for fewer levels and classes. The root has two children in 27 of the 36 fits; the 9 exceptions all have $\lambda \geq 0.02$ and the widest root has 6 children.

### 6.6 Independent embeddings and a cost that sees false joins

Sections 6.1 to 6.4 score the scheme on the graph it was fitted to. Three later runs of the survey's harness test whether its lead survives a different similarity, a corrected objective and a sweep of graph settings. All intervals in this section are paired item-bootstrap 95% percentile intervals over 1,000 draws.

**A second and third embedding model.** The corpus and holdout were re-embedded with two open models from other families, BAAI/bge-small-en-v1.5 and nomic-embed-text-v1.5, and graphs were built at $k = 10$, 20 and 50, giving 9 conditions with MiniLM (`bench/results/2026-10-10-embed`). The placements do not change; each is rescored on every graph. Table 9 counts the conditions in which the learned scheme leads the best other row of the same comparison (any surveyed scheme, gold or placed) and gives the tightest one.

In every comparison and every condition the learned scheme ranks first by skill, and both margins have intervals above zero. The tightest case is the held-out split on bge-small-k10, a skill margin of 0.155 [0.122, 0.189]. The edge does shrink off the fit graph. On the held-out split at $k = 20$ the skill margin is 0.190 on MiniLM, 0.166 on bge-small and 0.167 on nomic, and held-out skill falls from 0.922 to 0.901 and 0.891. Refitting on each of the other two models gives the mirror image (Table 10): each fit scores best on its own model's graph, by 0.926 against 0.904 held-out skill on average. That difference is home advantage, and it is a small part of the margin.

**Table 11.** Learned scheme minus the best other run under $J'$ (Dasgupta cost $C$ plus a specificity-weighted false-join term plus a navigation term, as defined in the survey) and under $C + \mathrm{FJ}$ (outside-mass form, both at $\beta = 1$). Negative means the learned scheme is ahead. Books and holdout are inserted as new works; the papers are the fitted sample.

| universe | model | k | best other by J' | J' learned - best [95% CI] | best other by C + FJ | C + FJ learned - best [95% CI] |
|---|---|---|---|---|---|---|
| 3,000 books | minilm | 10 | lcc-gold | -0.016 [-0.043, +0.007] | lcc-gold | +0.140 [+0.111, +0.167] |
| 3,000 books | minilm | 20 | ddc-section | -0.105 [-0.129, -0.083] | lcc-full-gold | +0.045 [+0.020, +0.068] |
| 3,000 books | minilm | 50 | ddc-d2 | -0.152 [-0.171, -0.134] | lcc-gold | -0.037 [-0.059, -0.016] |
| 3,000 books | bge-small | 10 | lcc-gold | +0.018 [-0.008, +0.040] | lcc-gold | +0.183 [+0.155, +0.209] |
| 3,000 books | bge-small | 20 | ddc-section | -0.052 [-0.075, -0.032] | lcc-full-gold | +0.092 [+0.068, +0.115] |
| 3,000 books | bge-small | 50 | ddc-d2 | -0.097 [-0.116, -0.079] | ddc-gold | +0.015 [-0.007, +0.038] |
| 3,000 books | nomic | 10 | lcc-gold | +0.036 [+0.010, +0.060] | lcc-gold | +0.195 [+0.168, +0.223] |
| 3,000 books | nomic | 20 | ddc-section | -0.058 [-0.082, -0.036] | lcc-full-gold | +0.098 [+0.072, +0.123] |
| 3,000 books | nomic | 50 | ddc-section | -0.108 [-0.129, -0.090] | lcc-gold | +0.010 [-0.012, +0.033] |
| 10,000 papers (in sample) | minilm | 20 | fdc-arxiv-nested | -0.136 [-0.145, -0.128] | arxiv-gold | -0.041 [-0.051, -0.031] |
| 10,000 papers (in sample) | bge-small | 20 | fdc-arxiv-nested | -0.115 [-0.122, -0.106] | arxiv-gold | -0.046 [-0.056, -0.036] |
| 10,000 papers (in sample) | nomic | 20 | fdc-arxiv-nested | -0.116 [-0.124, -0.108] | arxiv-gold | -0.036 [-0.047, -0.027] |
| 500 holdout books (title only) | minilm | 20 | lcc-pipeline | -0.264 [-0.309, -0.221] | lcc-full-gold | -0.249 [-0.306, -0.192] |
| 500 holdout books (title only) | bge-small | 20 | fdc-pipeline | -0.181 [-0.226, -0.137] | fdc-work | -0.110 [-0.168, -0.053] |
| 500 holdout books (title only) | nomic | 20 | fdc-work | -0.184 [-0.235, -0.126] | fdc-work | -0.134 [-0.198, -0.066] |

**Table 10.** Held-out skill of each learned fit (rows) on each model's $k = 20$ papers graph (columns).

| learned fit | scored on minilm | scored on bge-small | scored on nomic |
|---|---|---|---|
| fitted on minilm | 0.922 | 0.901 | 0.891 |
| fitted on bge-small | 0.913 | 0.938 | 0.909 |
| fitted on nomic | 0.899 | 0.912 | 0.918 |

This test removes the circularity of the *scoring* graph only. The held-out and holdout works are still filed by their MiniLM vectors, and Section 8.1 quantifies what the refits show about it.

**A cost that sees false joins.** Dasgupta cost charges for separating similar works and nothing for joining dissimilar ones. The survey adds a false-join term over a seeded sample of dissimilar non-edge pairs and validates the corrected objective $J'$ against a blind judge; the definitions and the validation are there. It has a specificity form ($J' = C + \mathrm{FJs} + \lambda' N$, with the survey's weight $\lambda'$, not the 0.005 this scheme was fitted with) and an outside-mass form $C + \mathrm{FJ}$, which is one minus the normalised dissimilarity objective of Cohen-Addad et al. [22]. For the learned scheme placed by the lexical classifier, the judge's agreement (Cohen's $\kappa$ [30]) does not change with the term ($\kappa$ 0.33 to 0.30, change -0.03 [-0.20, +0.14]): the judge rarely finds a false join in it.

Table 11 compares the learned scheme, with its own placements, against the best non-learned run under each form, on the graphs of the false-join rescoring (`bench/results/2026-10-11-falsejoin`). "Best other" is chosen by point value among all non-learned runs of that universe, including gold Dewey and LCC numbers on the books, before the interval is computed.

On the papers and the title-only holdout, the learned scheme leads under both forms in every graph (3 of 3 and 3 of 3 under $J'$). On the 3,000 books it does not hold. Under $J'$ it leads in 6 of 9 graphs, ties in 2 and is behind in 1; the ties and the loss are against LCC's librarian numbers on the $k = 10$ graphs. Under the outside-mass form it is behind in 6 of 9, ties in 2 and leads in 1. The outside-mass form charges a shallow join almost as much as a deep one, so it rewards a wide top level (`bench/results/2026-10-11-falsejoin/README.md`). The learned scheme has a binary root (Section 8.3), so every dissimilar pair that lands in the same half of it pays at least half the maximum.

**The 27-setting book sweep.** The survey repeated the book comparison in 27 settings: three models, three values of $k$ and three graph thresholds (`bench/results/2026-10-11-book-stability`). The claim that the learned scheme beats Dewey sections under $J'$ has the claimed sign in 19 of 27 settings and is significantly reversed in 6. By that run's own rule the claim is not stable.

**Verdict.** The learned scheme's edge survives an independent embedding. On arXiv papers it also survives the false-join correction. On library books, which it was not fitted to, it does not survive: under the corrected objective its lead over librarian-assigned Dewey and LCC numbers depends on the graph setting, and under the outside-mass form it is usually behind.

## 7 DISCUSSION

**What the prototype shows.** Under a stated objective, a fitted order-preserving scheme with 110 hand-labelled classes scores well below every published scheme on that objective with its own placements, in sample and on a held-out split. Through the same lexical classifier as the published schemes it is still below every published scheme's placed row (0.3794 against 0.6414 for the best), but above arXiv's own author-chosen categories (0.3333). Most of that gap is expected: the published schemes were not built for this corpus or this similarity, and several are domain schemes on foreign ground. The useful result is not the ranking but the reference point (the survey ranks the hand-built schemes). A hand-built scheme can now be described as "this far from a fitted scheme with the same number of levels on the same graph", and the cost of each constraint (order preservation, the depth bound, stable insertion) can be measured rather than argued.

**What survives the checks.** The lead holds when the scoring graph comes from another embedding model, less a home advantage (Section 6.6 and Section 8.1). It does not hold on books once false joins are charged. A fitted scheme that is excellent on its own kind of material and only level with librarian numbers on books is a reference point, not a replacement.

**The price of order preservation is not yet measured.** The DP collapse and the local search work only within one fixed shelf order, and the fitted tree keeps 97.6% of the average-linkage revenue. That loss mixes two effects: collapsing a binary tree to a bounded-shape class tree, and doing so without reordering works. Any binary tree has a compatible leaf order, so the order constraint costs nothing before the collapse; whether it costs much after it needs a bounded-shape fit without the order constraint, which we have not run. Whether the answer holds for a heterogeneous collection is open.

**Stability as a design constraint.** Separating batch fitting from incremental growth gives a guarantee a cataloguer can rely on (Section 4.2) at the price of drift: the inserted scheme is no longer $J$-optimal and its depth grows where new material concentrates. A periodic refit would restore optimality but break identifiers. A principled middle ground, an objective that charges for every renumbered work, is the open problem we would most like to see addressed (we found no prior formulation).

**Towards a general classification.** A general scheme needs a general corpus. The corpus here is papers from arXiv's top-level archives; a library's collection is mostly books, much of it in the humanities, arts and social sciences, with titles and subject headings rather than abstracts. The method transfers (it needs only a similarity graph and text for labels); the scheme does not, as the holdout shows.

## 8 LIMITATIONS

### 8.1 Embedding circularity: the held-out skill is optimistic

Fit, filing and scoring share one embedding. The graph is built from sentence-transformers/all-MiniLM-L6-v2 vectors, the fit minimises $J$ on that graph, and insertion files a new work at the nearest leaf centroid in the *same* vector space. A held-out work is therefore placed by exactly the geometry it is later scored on. This favours the learned rows in Table 4 and Table 6, and it is worst on the holdout, where both the graph and the insertion see only title embeddings. The published schemes' placed rows, and the learned scheme's own placed rows, use the lexical classifier and do not share this advantage. The paired intervals of Section 6 do not cover it: they resample works, not embeddings.

The embedding run (`bench/results/2026-10-10-embed/results.md`, "Cross-fit") measures part of it. It fits the scheme on each of three models (MiniLM, bge-small, nomic; all $k = 20$, automatic labels) and scores each fit on each model's graph. Held-out skill on the fit's own model's graph averages 0.926 over the three fits and on the two other models' graphs 0.904. For the MiniLM fit of this paper the figures are 0.922 on MiniLM and 0.896 on the other two (bge-small 0.901, nomic 0.891), a gap of 0.026. On the 500 books the gap is larger: skill averages 0.806 on the fit's own model's graph and 0.731 on the others, a gap of 0.075, and for the MiniLM fit it falls from 0.797 to 0.724. That run's own summary puts the home advantage at about 0.02 to 0.03 skill on the held-out split, which matches the 0.026 here; the book gap is about 2.9 times that.

These gaps bound the scoring-graph part only. In every cell of the cross-fit, held-out works are filed by the *fit's* embedding, so the off-diagonal cells still share fit and filing. No run files works with an embedding the scheme was not fitted in, and none scores the scheme on a similarity that is not an embedding (citation coupling, subject-heading overlap); the metric itself was checked against such signals in the survey, the scheme was not. We therefore read the held-out skill 0.922 as an upper estimate, whose size beyond the 0.026 above is unknown, and the claim that the learned scheme leads the existing ones as one that survives a change of scoring embedding (Section 6.6) and not yet a change of filing embedding.

### 8.2 Fitted on arXiv only

The scheme has seen only arXiv papers. Its top level splits computing from mathematics and the physical sciences; it has no class for the humanities, law, arts or literature until insertion creates one, and then a weak one. It is a general *method*, not yet a general classification. The arXiv sample itself is proportional to archive size with a floor per archive (`bench/corpus/README.md`), and computer science is its largest part: 39.7% of the papers have a `cs` primary category.

### 8.3 A binary root under this objective

The root has 2 children in the committed scheme and in 27 of 36 sweep fits. Dasgupta cost prefers a binary split near the top, where most cross-cluster weight sits, and the navigation term charges only per level, so it cannot prefer a wide first menu. Library schemes have wide top levels (LCC's first level is its set of lettered main classes [31]) for reasons of browsing and memorability that $J$ does not model. A top-level fan-out prior, or a navigation cost with a linear scan term (the harness's `nav_scan`), may change this; we have not tested either.

### 8.4 Weak automatic labels for inserted classes

Classes created by insertion are labelled from their works' most frequent title words. Table 7 shows the result: labels such as "Harry / potter / adventures" for the first class of books, and a chain of "(general)" core classes that repeat their parent's label down 9 levels. These are not usable as catalogue captions. The hand labels for fitted classes were written by an AI coding assistant from the label sheet, not by a cataloguer, and have not been evaluated by people, nor scored with any of the taxonomy-quality attributes in the literature [32].

### 8.5 Classifier coverage on book titles

The class vocabulary is mined from arXiv titles and abstracts, so the lexical classifier rarely matches book titles: 43 of 500 holdout books (8.6%) were placed, and the non-circular placed comparison on books is uninformative for this scheme until its vocabulary covers books.

### 8.6 Graph-hash non-reproducibility with near-tied edges

Rebuilding the published graphs from the pinned embedding weights (`model.onnx` sha256 `bbd7b466f6d5`...) on a different ONNX Runtime version gives edge weights that differ by floating-point noise and, in a few places, a different choice between near-tied neighbours, so the rebuilt graph's hash differs (`bench/results/2026-10-10-learned/REPRODUCE.md`). The fit uses only the published graph, so the scheme rebuilds byte for byte; the graph file, not the embedding pipeline, is the reproducible artefact. Insertion uses i8-quantized centroids, whose step is coarser than the noise.

### 8.7 Other limits

- Insertion does not keep the depth bound: holdout insertion produced classes 9 levels deep and a mean work depth of 8.79 in the harness's tree (Table 6).
- One corpus to fit. Table 1 to Table 8 are point values on the fit graph; the paired intervals are in Section 6.6. A permutation null for the learned rows has not been run.
- The false-join term used in Section 6.6 is validated in the survey against one LLM judge; its limits are stated there.
- No user study: $N$ is a model of navigation cost [24], [25], not a measurement. User studies of menu depth and breadth disagree on the best shape: speed and accuracy improved with breadth in one [33], others found an optimum at moderate breadth [34], [35], and none of these was run on a library classification, so they do not calibrate $N$.
- The comparison mixes gold rows (a source's own labels), placed rows (a classifier's output) and own rows (fitted or inserted); they answer different questions and should not be read as one ranking.

## 9 RELEASE AS CC0 OPEN DATA

`schemes/learned/learned.json` (`scheme:learned@2026-10-10`, sha256 `379f746e99ad`...), its hand-written labels (`labels.tsv`), and the leaf centroids and quantized holdout vectors in `schemes/learned/data/` are dedicated to the public domain under CC0 1.0 [36]. The builder code is MIT OR Apache-2.0 like the rest of the repository. CC0 is chosen so that the scheme, unlike DDC, can be copied into catalogues, crosswalks and derived schemes without conditions, and so that it combines with other CC0 sources: arXiv metadata [37] and Open Library's catalogue data, which Open Library describes as dedicated to the public domain and asks contributors to share under CC0 1.0 [38]. The scheme validates against the repository's scheme schema (`shelfmark scheme validate --path schemes/learned/learned.json`), and `sources.lock.json` records the sha256 of every input. The scheme identifiers are versioned (`learned@2026-10-10`, `learned@2026-10-10.ins500` for the holdout insertion), and the stability guarantee of Section 4 is part of the release: a future version that grows by insertion keeps every code of this one.

Repository: https://github.com/davidawad/shelfmark (the scheme is in `schemes/learned/`).

## 10 REPRODUCIBILITY

- **Inputs.** Corpus, holdout, the published $k = 20$ graphs and the scheme are committed with their sha256 values; the graphs were built with sentence-transformers/all-MiniLM-L6-v2 weights revision `d13954661f83`.
- **Fit.** `cargo run --release --manifest-path schemes/learned/Cargo.toml -- build ...` (commands in `schemes/learned/README.md`) rebuilds `learned.json` byte for byte from committed files in seconds; the test `rebuild_reproduces_the_committed_scheme_and_matches_the_harness` enforces this, and `sources_lock_and_registry_match_the_committed_files` checks the input hashes.
- **Benchmark.** `bench/results/2026-10-10-learned/REPRODUCE.md` gives every command, including the survey's placement rerun, embedding the vectors for the centroids, and `crates/bench/scripts/learned_benchmark.py`. The results record the tool commit (`6b9ff9de9a66`) and binary hashes.
- **Determinism.** The fit is seeded (20261010) and single-threaded; insertion is deterministic (tested). The one non-reproducible step is regenerating the graph from text (Section 8.6).
- **This paper.** `python3 paper/paper2/scripts/make_tables.py` regenerates `generated.typ` from the JSON; `--check` fails if it is stale. The script cross-checks $J$ against its two terms for every row, the placed rows against the survey's rerun, and the fit report against the harness.
- **Machine-checked claims.** The identifier-stability guarantee (G1 to G4), the contiguity of classes over a fixed shelf order, the exactness of the collapse dynamic program, and the exact optimal order-preserving tree for a fixed shelf order are proved in Lean 4 in the repository's `proofs/` directory, with no `sorry` and no axioms beyond Lean's standard three. The proofs cover abstract models of the algorithms, not the Rust code line by line.

## ACKNOWLEDGEMENTS

The author thanks Hetzner for offering an API for autonomous server provisioning; DeepSeek for releasing open-source models; and Anthropic for Claude. Both DeepSeek and Claude assisted with the analysis, code and drafting. The validity labels were produced by an Anthropic model acting as the judge.

## REFERENCES

[1] D. Awad, “A Survey of Open Knowledge Classification Schemes,” 2026.

[2] S. Dasgupta, "A cost function for similarity-based hierarchical clustering," in *Proceedings of the 48th Annual ACM Symposium on Theory of Computing (STOC)*, 2016, pp. 118–127. doi: 10.1145/2897518.2897527.

[3] T. C. Hu and A. C. Tucker, "Optimal Computer Search Trees and Variable-Length Alphabetical Codes," *SIAM Journal on Applied Mathematics*, vol. 21, no. 4, pp. 514–532, 1971, doi: 10.1137/0121057.

[4] J. P. Comaromi, "Conception and development of the Dewey Decimal Classification," *Knowledge Organization*, vol. 3, no. 1, pp. 11–15, 1976, doi: 10.5771/0943-7444-1976-1-11.

[5] B. Hjørland, "Dewey Decimal Classification (DDC)," *Em Questão*, vol. 31, p. e–145068, 2025, doi: 10.1590/1808-5245.31.145068.

[6] C. Fraser, "Mathematics in Library and Review Classification Systems: An Historical Overview," *Knowledge Organization*, vol. 47, no. 4, pp. 334–356, 2020, doi: 10.5771/0943-7444-2020-4-334.

[7] L. Waltman and N. J. van Eck, "A new methodology for constructing a publication-level classification system of science," *Journal of the American Society for Information Science and Technology*, vol. 63, no. 12, pp. 2378–2392, 2012, doi: 10.1002/asi.22748.

[8] V. A. Traag, L. Waltman, and N. J. van Eck, "From Louvain to Leiden: guaranteeing well-connected communities," *Scientific Reports*, vol. 9, p. 5233, 2019, doi: 10.1038/s41598-019-41695-z.

[9] J. Priem, H. Piwowar, and R. Orr, "OpenAlex: A fully-open index of scholarly works, authors, venues, institutions, and concepts." [Online]. Available: https://arxiv.org/abs/2205.01833

[10] OpenAlex, "Topics." [Online]. Available: https://help.openalex.org/hc/en-us/articles/24736129405719-Topics

[11] C. Zhang *et al.*, "TaxoGen: Unsupervised Topic Taxonomy Construction by Adaptive Term Embedding and Clustering," in *Proceedings of the 24th ACM SIGKDD International Conference on Knowledge Discovery & Data Mining (KDD)*, 2018, pp. 2701–2709. doi: 10.1145/3219819.3220064.

[12] Q. Zeng *et al.*, "Chain-of-Layer: Iteratively Prompting Large Language Models for Taxonomy Induction from Limited Examples," in *Proceedings of the 33rd ACM International Conference on Information and Knowledge Management (CIKM)*, 2024, pp. 3093–3102. doi: 10.1145/3627673.3679608.

[13] M. Wan *et al.*, "TnT-LLM: Text Mining at Scale with Large Language Models," in *Proceedings of the 30th ACM SIGKDD Conference on Knowledge Discovery and Data Mining (KDD)*, 2024, pp. 5836–5847. doi: 10.1145/3637528.3671647.

[14] D. E. Knuth, "Optimum binary search trees," *Acta Informatica*, vol. 1, no. 1, pp. 14–25, 1971, doi: 10.1007/BF00264289.

[15] D. A. Huffman, "A Method for the Construction of Minimum-Redundancy Codes," *Proceedings of the IRE*, vol. 40, no. 9, pp. 1098–1101, 1952, doi: 10.1109/JRPROC.1952.273898.

[16] J. Shen, Z. Shen, C. Xiong, C. Wang, K. Wang, and J. Han, "TaxoExpan: Self-supervised Taxonomy Expansion with Position-Enhanced Graph Neural Network," in *Proceedings of The Web Conference 2020*, 2020, pp. 486–497. doi: 10.1145/3366423.3380132.

[17] N. Reimers and I. Gurevych, "Sentence-BERT: Sentence Embeddings using Siamese BERT-Networks," in *Proceedings of the 2019 Conference on Empirical Methods in Natural Language Processing and the 9th International Joint Conference on Natural Language Processing (EMNLP-IJCNLP)*, 2019, pp. 3980–3990. doi: 10.18653/v1/D19-1410.

[18] W. Wang, F. Wei, L. Dong, H. Bao, N. Yang, and M. Zhou, "MiniLM: Deep Self-Attention Distillation for Task-Agnostic Compression of Pre-Trained Transformers." [Online]. Available: https://arxiv.org/abs/2002.10957

[19] sentence-transformers, "all-MiniLM-L6-v2." [Online]. Available: https://huggingface.co/sentence-transformers/all-MiniLM-L6-v2

[20] B. Moseley and J. R. Wang, "Approximation Bounds for Hierarchical Clustering: Average Linkage, Bisecting K-means, and Local Search," *Journal of Machine Learning Research*, vol. 24, no. 1, pp. 1–36, 2023, [Online]. Available: https://jmlr.org/papers/v24/18-080.html

[21] M. Charikar and V. Chatziafratis, "Approximate Hierarchical Clustering via Sparsest Cut and Spreading Metrics," in *Proceedings of the Twenty-Eighth Annual ACM-SIAM Symposium on Discrete Algorithms (SODA)*, 2017, pp. 841–854. doi: 10.1137/1.9781611974782.53.

[22] V. Cohen-Addad, V. Kanade, F. Mallmann-Trenn, and C. Mathieu, "Hierarchical Clustering: Objective Functions and Algorithms," *Journal of the ACM*, vol. 66, no. 4, pp. 1–42, 2019, doi: 10.1145/3321386.

[23] I. Chami, A. Gu, V. Chatziafratis, and C. Ré, "From Trees to Continuous Embeddings and Back: Hyperbolic Hierarchical Clustering," in *Advances in Neural Information Processing Systems 33 (NeurIPS 2020)*, 2020, pp. 15065–15076. [Online]. Available: https://papers.nips.cc/paper_files/paper/2020/hash/ac10ec1ace51b2d973cd87973a98d3ab-Abstract.html

[24] W. E. Hick, "On the Rate of Gain of Information," *Quarterly Journal of Experimental Psychology*, vol. 4, no. 1, pp. 11–26, 1952, doi: 10.1080/17470215208416600.

[25] R. Hyman, "Stimulus information as a determinant of reaction time," *Journal of Experimental Psychology*, vol. 45, no. 3, pp. 188–196, 1953, doi: 10.1037/h0056940.

[26] J. M. Kleinberg, "An Impossibility Theorem for Clustering," in *Advances in Neural Information Processing Systems 15 (NIPS 2002)*, 2002. [Online]. Available: https://papers.nips.cc/paper_files/paper/2002/hash/43e4e6a6f341e00671e123714de019a8-Abstract.html

[27] G. Carlsson and F. Mémoli, "Characterization, Stability and Convergence of Hierarchical Clustering Methods," *Journal of Machine Learning Research*, vol. 11, no. 47, pp. 1425–1470, 2010, [Online]. Available: https://www.jmlr.org/papers/v11/carlsson10a.html

[28] D. Boley, "Principal Direction Divisive Partitioning," *Data Mining and Knowledge Discovery*, vol. 2, no. 4, pp. 325–344, 1998, doi: 10.1023/A:1009740529316.

[29] M. Grootendorst, "BERTopic: Neural topic modeling with a class-based TF-IDF procedure." [Online]. Available: https://arxiv.org/abs/2203.05794

[30] J. Cohen, "A Coefficient of Agreement for Nominal Scales," *Educational and Psychological Measurement*, vol. 20, no. 1, pp. 37–46, 1960, doi: 10.1177/001316446002000104.

[31] Library of Congress, "Library of Congress Classification PDF Files." [Online]. Available: https://www.loc.gov/aba/publications/FreeLCC/freelcc.html

[32] M. Unterkalmsteiner and W. Abdeen, "A compendium and evaluation of taxonomy quality attributes," *Expert Systems*, vol. 40, no. 1, 2023, doi: 10.1111/exsy.13098.

[33] K. Snowberry, S. R. Parkinson, and N. Sisson, "Computer display menus," *Ergonomics*, vol. 26, no. 7, pp. 699–712, 1983, doi: 10.1080/00140138308963390.

[34] D. P. Miller, "The Depth/Breadth Tradeoff in Hierarchical Computer Menus," *Proceedings of the Human Factors Society Annual Meeting*, vol. 25, no. 1, pp. 296–300, 1981, doi: 10.1177/107118138102500179.

[35] K. Larson and M. Czerwinski, "Web page design," in *Proceedings of the SIGCHI conference on Human factors in computing systems - CHI '98*, ACM Press, 1998, pp. 25–32. doi: 10.1145/274644.274649.

[36] Creative Commons, "CC0 1.0 Universal Public Domain Dedication." [Online]. Available: https://creativecommons.org/publicdomain/zero/1.0/

[37] arXiv, “arXiv License Information.” [Online]. Available: https://info.arxiv.org/help/license/index.html
[38] Open Library, “Using Open Library Data.” [Online]. Available: https://openlibrary.org/help/faq/using